\documentclass[preprint]{revtex4}
\newcommand{\ba}{\begin{eqnarray}}
\newcommand{\ea}{\end{eqnarray}}

\usepackage{graphics}

\begin{document} 

\title{Boundary conditions for the electronic structure of finite-extent,
embedded semiconductor nanostructures} 
\author{Seungwon Lee, Fabiano Oyafuso, Paul von Allmen, and Gerhard Klimeck} 
\affiliation{Jet Propulsion Laboratory, California
Institute of Technology, Pasadena, California 91109} 
\date{\today}

\begin{abstract} 
The modeling of finite-extent semiconductor nanostructures
that are embedded in a host material requires a proper boundary treatment
for a finite simulation domain. For the study of a self-assembled InAs
dot embedded in GaAs, three kinds of boundary conditions are examined within
the empirical tight-binding model: (i) the periodic boundary condition, (ii)
raising the orbital energies of surface atoms, and (iii) raising the energies
of dangling bonds at the surface. The periodic boundary condition requires a
smooth boundary and consequently a larger GaAs buffer than the two non-periodic
boundary conditions. Between the non-periodic conditions, the dangling-bond
energy shift is more numerically efficient than the orbital-energy shift, 
in terms of the elimination of non-physical surface states in the energy region of interest
for interior states. A dangling-bond energy shift larger than 5~eV efficiently 
eliminates all of the surface states and leads to interior states 
that are highly insensitive to the choice of the energy shift.  
\end{abstract}

\maketitle 

\section{Introduction}

The representation of a semiconductor heterostructure by an atomistic model ultimately requires the
introduction of a limited simulation domain, of which the surface needs to be treated with 
a specific boundary condition (BC). 
If the surface of the simulation domain is selected far enough from the central feature of interest, 
periodic BCs can be used and the simulation domain is effectively repeated infinitely.  
However, for electronic devices with non-periodic external potentials
or for structures with irregular surfaces, the periodic BCs are not a natural choice.  
If the simulation-domain surface is within the material bulk region, 
a truly open BC or perfectly absorbing BC would be the best solution, 
as it does not introduce an artificial periodicity and would
enable the simulation of carrier injection or transport.\cite{klimeck_apl, mamaluy}  
However such a BC requires the inversion of a full matrix that is of the order of the number of
atoms on the open surface. Therefore, the open BC can only be applied to relatively
small open surfaces.  

Another choice in representing a finite simulation domain is the abrupt termination of the
simulation domain with a hard-wall BC.  Such abrupt termination in the atomistic basis
set results in the creation of dangling bonds.  The dangling bonds will form surface states
(of the order of the number of exposed atoms), that typically cover a broad energy range and
often litter the central energy region of the fundamental band gap.  The separation of the
artificially introduced surface states from the desired centrally confined states is numerically
expensive, as the computation time and required memory increase with the number of computed
eigenvalues and eigenvectors and as the separation would demand the computation of eigenvectors. Many
relevant quantum dot calculations only require the computation of eigenvalues,\cite{klimeck-nemo3d}
while the computation of the eigenvectors at least doubles the computation time and the required memory
scales with the number of computed eigenvectors. To address the problem of artificially introduced
surface states, this paper examines two modified hard-wall BCs and discusses their merits relative 
to each other and to the more standard periodic BC.

Typical quantum-dot and heterostructure devices are based on the concept of confining electron and
hole states into a spatial domain. The confinement is typically achieved by surrounding a core
semiconductor by a buffer semiconductor of larger band gap. The practical question now
arises of how large of a buffer region must be included in the explicit simulation domain. In
systems of strain-induced self-assembled quantum dots the strain fields may extend out from the
central device region for tens of nanometers,\cite{oyafuso} while the quantum states of interest
extend only over a few atomic monolayers into the buffer.  The lattice distortion due to strain must
therefore be computed in a large simulation domain, while the desired quantum confined states may
only need to be computed in a relatively small simulation domain. The hard wall BCs 
considered in this paper enable the strain and electronic structure simulations 
to be performed with two different simulation domains.  
This paper demonstrates that the inclusion of a realistically
large buffer is essential to capture the effects of strain, while the subsequent electronic
structure calculation can then be performed with a significantly smaller, strain distorted simulation
domain which resolves the confined quantum states of interest. 
The reduction of the simulation domain for the electronic structure calculation substantially 
lessens the computational requirements since the dimension of the Hamiltonian grows linearly 
with the number of atoms included in the model. 

The proper BC for a reduced buffer should efficiently eliminate 
all non-physical surface states and at the same time should minimally affect physical 
interior states. In previous work, two types of BC have been considered for 
the atomistic modeling of embedded nanostructures.\cite{oyafuso, wang-kim-zunger}
In the first BC, the orbital energy of the surface atoms is
raised by a specific amount.\cite{oyafuso} The value of energy shift is determined
empirically by requiring that no state resides in the energy
gap.\cite{oyafuso} We will show that this method is unpredictable and numerically 
less efficient than the new BC proposed in this work. 
The second BC found in the literature is the periodic BC with 
a truncated buffer.\cite{wang-kim-zunger} 
We also find this method inefficient in eliminating spurious states formed 
in the energy gap region as it requires either a relatively larger buffer or
an unphysical, empirical adjustment to atomic positions near the boundary for a small buffer.  
In the present work, we propose a new BC that is to raise the energy of dangling bonds.
We compare the proposed BC with the two previously employed BCs and 
demonstrate the efficiency and reliability of the new BC. 
The three boundary conditions are applied to the study of the electronic
structure of a self-assembled InAs quantum dot embedded in a GaAs buffer
in the framework of the empirical tight-binding model.
The efficiency and reliability of the BCs are measured by 
the elimination of non-physical surface states, 
the number of iterations in the Lanczos eigenvalue solver, and 
the reduction of the buffer size required for interior-state energy convergence. 

\section{Boundary Conditions} \label{sec:boundary}

The first boundary condition (BC~I) considered is to raise the orbital energies
of surface atoms. This method discourages electrons from populating the
surface-atom orbitals. However, this treatment does not differentiate details
of the surface atoms such as the number and direction of their dangling bonds. 
As a refinement, a second boundary condition (BC~II) is introduced:  
raising the energy of the dangling bond for the
surface atoms. Within this method, the connected-bond energy of the surface atoms
is kept unchanged and hence there is no extra penalty for electrons to occupy
the connected bonds of surface atoms. Since the motivation of the surface
energy shift in BC~I and II is to remove non-physical surface states from
the energy region of interest, lowering the surface energies will have 
the same outcome as raising the surface energies. 

Both BC~I and II are closed boundary conditions as opposed to a periodic condition
that is the third boundary condition (BC~III) considered in this work.
In principle, this boundary condition is applicable only if the system is composed
of a unit cell periodically repeated.  However, the periodic boundary
condition is widely used not only for periodic systems but also for
systems with non-periodic perturbations such as alloy disorder, defects,
impurities, and even surfaces.  For systems with such non-periodic
perturbations, the unit cell known as the supercell should be large
enough to accommodate the non-periodic perturbations.  In nanostructure modeling, the
supercell can be as large as the whole size of the nanostructures. For
instance, the nanostructure composed of a quantum dot and a surrounding buffer 
has no inherent periodicity, with a long-ranged strain field that extends up to tens
of nanometers.\cite{oyafuso} The periodic boundary condition is therefore
examined for its appropriateness and efficiency in modeling these
nanostructures. 

These three boundary conditions are implemented in the framework of
the orthogonal nearest-neighbor tight-binding model. In this model, 
the effective Hamiltonian is expressed as the sum of the couplings 
between atomic basis orbitals $|i,\gamma\rangle$:
\ba
H_0=\sum_{i\gamma} \epsilon_{\gamma}|i, \gamma\rangle \langle i,\gamma|
  + \sum_{i \neq i'\gamma\gamma'}t_{ii'\gamma\gamma'} |i,\gamma\rangle \langle
i',\gamma'|,
\ea 
where indices $i$ and $\gamma$ denote an atomic site and an orbital type. 
Parameter $\epsilon$ represents the energy of the basis orbital, 
and $t$ accounts for the coupling between basis orbitals centered at 
nearest-neighbor atomic sites.

In BC~I, the Hamiltonian block matrix for a surface atom with basis set 
\{$|s\rangle, |p_x\rangle,
|p_y\rangle, |p_z\rangle$\} is given by
\ba
\left[
  \begin{array}{ccccc}
	\epsilon_s +\delta_s & 0 & 0 & 0  \\
	0 & \epsilon_p +\delta_p & 0 & 0  \\
        0 & 0 & \epsilon_p + \delta_p & 0 \\
        0 & 0 & 0 & \epsilon_p + \delta_p 
  \end{array}
\right],
\label{eq:bc1}
\ea
where $\delta_{\gamma}$ is the energy shift for the orbital $\gamma$
on a surface atom. A different energy shift can be chosen for each basis orbital.

For BC~II,  the basis set of the Hamiltonian is first changed 
from set \{$|s\rangle, |p_x\rangle, |p_y\rangle, |p_z\rangle$\} to the set  
of $sp^3$ hybridized orbitals that are aligned along the bond directions.
In the zinc-blende structure, the $sp^3$ hybridized orbitals are 
given by\cite{bond_direction}
\ba
\begin{array}{c}
|sp^3_a\rangle = \frac{1}{2}(|s\rangle +|p_x\rangle +|p_y\rangle +|p_z\rangle),\\ 
|sp^3_b\rangle = \frac{1}{2}(|s\rangle +|p_x\rangle -|p_y\rangle -|p_z\rangle),\\
|sp^3_c\rangle = \frac{1}{2}(|s\rangle -|p_x\rangle +|p_y\rangle -|p_z\rangle),\\
|sp^3_d\rangle = \frac{1}{2}(|s\rangle -|p_x\rangle -|p_y\rangle +|p_z\rangle).
\end{array}
\ea

The energy of a hybridized orbital is raised by $\delta_{sp^3}$ if the
orbital is along the dangling bond direction. For instance, if the surface
atom has dangling bonds along $|sp^3_a\rangle$ and $|sp^3_c\rangle$
directions, the Hamiltonian block matrix for the surface atom in the basis set
$\{ |sp^3_a\rangle, |sp^3_b\rangle, |sp^3_c\rangle, |sp^3_d\rangle\}$ is
given by
\ba
\left[
\begin{array}{ccccc}
a+\delta_{sp^3} & b & b & b  \\
b & a & b & b \\
b & b & a+\delta_{sp^3} & b  \\
b & b & b & a 
\end{array}
\right],
\ea
where $a=\epsilon_s/4+3\epsilon_p/4$ and $b=\epsilon_s/4-\epsilon_p/4$.  

Finally, the Hamiltonian is transformed back into the original basis set of
$\{|s\rangle, |p_x\rangle, |p_y\rangle, |p_z\rangle\}$. The final
Hamiltonian block matrix for the surface atom becomes
\ba
\left[
\begin{array}{ccccc}
\epsilon_s+\frac{\delta_{sp^3}}{2} & 0 & \frac{\delta_{sp^3}}{2} & 0  \\
0 & \epsilon_p+\frac{\delta_{sp^3}}{2} & 0 & \frac{\delta_{sp^3}}{2}  \\
\frac{\delta_{sp^3}}{2} & 0 & \epsilon_p+\frac{\delta_{sp^3}}{2} & 0  \\
0 & \frac{\delta_{sp^3}}{2} & 0 & \epsilon_p+\frac{\delta_{sp^3}}{2} 
\end{array}
\right].
\ea

In comparison with Eq.~(\ref{eq:bc1}), this block matrix contains nonzero
off-diagonal elements.  Furthermore, the shift of the diagonal element is
proportional to the number of dangling bonds.  If the surface atom has $n$
dangling bonds, the energy shift of the diagonal elements is given by
$n\delta_{sp^3}/4$. This shows that BC~II distinguishes among
surface atoms with a different number of dangling bonds. 
It is important to note that BC~II becomes identical to BC~I when 
the energies of all the four $sp^3$ hybridized orbitals are raised by the same amount. 
Therefore, BC~I can be interpreted as the boundary condition that 
truncates the dangling bonds as well as the bonds connected to interior atoms. 

To some degree, BC~II mimics the physical passivation of dangling bonds with
other atoms such as hydrogen and oxygen. Experimentally, silicon surfaces are
usually passivated by hydrogen to improve the conductivity. The hydrogen forms
bonding and anti-bonding states with the dangling bonds of Si at the surface.
For example, the energies of the bonding and anti-bonding states of SiH$_4$ are
about 18 eV and 5 eV below the valence band edge of bulk Si,
respectively.\cite{cardona} Therefore, hydrogen passivation efficiently removes
surface states localized in dangling bonds. In connection with this mechanism,
BC~II can be interpreted as the approximate formation 
of the bonding and antibonding states between a dangling
bond and vacuum at an energy determined by
$\delta_{sp^3}$.\cite{surface_passivation}

Although BC~I and II can be also applied to excited orbitals such as $d$ and $s^*$, 
it is unnecessary to shift the energies of the excited orbitals for surface atoms. 
The atomic energies of the excited orbitals (typically 10--20~eV) 
are larger than the energy gap, which is typically 0--5~eV.\cite{jancu}   
Furthermore, the bonding states between the excited orbital and the $s/p$ orbital
are shifted up by the energy shift of the $s/p$ orbitals.   
Therefore, the unmodified excited orbitals of surface atoms do not lead to 
surface states in the middle of the energy gap.      

Implementing BC~I and II requires a proper choice for the
energy shift of the surface atoms. The energy shift should be high enough to
discourage electrons from occupying the surface atom orbitals and consequently
to eliminate all non-physical surface states in the middle of the gap.  
The diagonal elements of the tight-binding Hamiltonian give a guide to 
the required energy shift. The diagonal elements range from 0.6 eV to 20 eV.
The sensitivity of the electronic structure to different energy shifts $\delta_s$,
$\delta_p$, and $\delta_{sp^3}$ is discussed in Section~\ref{sec:shift}.

Finally for BC~III (the periodic boundary condition), 
every surface atom is connected with another surface atom on the 
opposite side of the supercell. Consequently,  
the coupling between surface atoms from the two sides is added to 
the original Hamiltonian:
\ba
H_{periodic} = H_0 + \sum_{\langle jk \rangle \gamma\gamma'}
t_{jk\gamma\gamma'}|j,\gamma\rangle \langle k,\gamma'|, 
\ea
where $\langle jk \rangle$ denotes all the new pairs of neighbors due to the
periodic boundary condition. The diagonal block matrix of the
Hamiltonian for surface atoms is unchanged in the periodic boundary condition 
as opposed to BC~I and II. 

\section{Nanostructure Modeling}

The three boundary conditions are applied to the study of the electronic
structure of a self-assembled InAs quantum dot embedded in a GaAs buffer. The
modeled dot is lens shaped with diameter 15 nm and height 6 nm, similar
to experimentally available dots.\cite{schmidt-petroff, garcia-petroff}.  The
appropriate size for the GaAs buffer depends on the type of calculation.
For strain-profile calculations, the buffer thickness should be at least 
as large as the dot size since the strain field is long-ranged, 
while for electronic-structure calculations 
the buffer thickness can be smaller than the dot size because
bound electron states are effectively confined inside the dot.\cite{oyafuso} In
this work, a 15~nm thick buffer is used for the strain-profile
calculation, and a reduced buffer with thickness 1 -- 5~nm is used for the electronic
structure calculation with the atomic positions given by the larger strain
calculation.  The equilibrium atomic positions are calculated by minimizing the
strain energy using an atomistic valence-force-field
model.\cite{klimeck-nemo3d,keating,pryor}
The necessity of a large buffer size for the strain calculation 
and the long-range effect of the strain on the electronic structure
are discussed in Section~\ref{sec:strain}.
Under the saturated strain profile obtained with a sufficiently large buffer, 
the quantitative effect of the reduced buffer
size on the electronic structure is examined in Section~\ref{sec:buffer}. 

The tight-binding Hamiltonian for the InAs dot and the GaAs buffer is
constructed based on atomic $sp^3d^5s*$ orbitals.
The Hamiltonian matrix elements are obtained by fitting to experimental bulk
band structure parameters with a genetic optimization 
algorithm.\cite{klimeck-nemo3d, boykin_strain}
To take into account the effect of the displacements of atoms from the
unstrained crystal positions, the atomic energies (the diagonal elements of the
Hamiltonian) are adjusted by a linear correction within the
L\"owdin orthogonalization procedure.\cite{boykin_strain,lowdin} The coupling
parameters between nearest-neighbor orbitals (the off-diagonal elements of the
Hamiltonian) are also modified according to the generalized
Harrison $d^{-2}$ scaling law and Slater-Koster direction-cosine
rules.\cite{harrison, slater-koster}

The eigenvalues of the tight-binding Hamiltonian is obtained with 
the Lanczos algorithm,\cite{treffethen} which is a commonly used
iterative eigenvalue solver for large-dimensional, sparse, Hermitian matrices,
as is the case for our tight-binding Hamiltonian. At each Lanczos iteration,
the matrix is projected into a lower-dimensional subspace known as the
Krylov subspace.  The reduced matrix is tridiagonal and its eigenvalues
approximate those of the original matrix as the size of the Krylov
subspace grows. 

\section{Results and Discussion}

\begin{figure}[t]\scalebox{0.65}{\includegraphics*{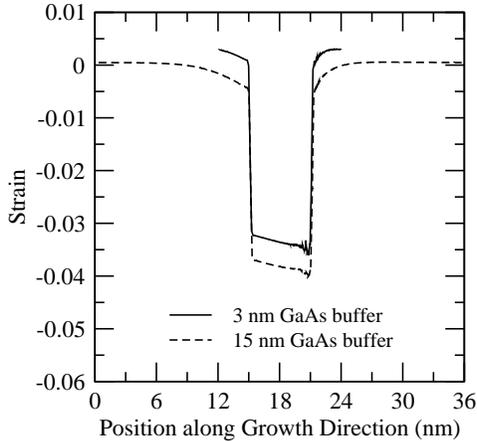}}
\caption{\label{fig:strain_profile} 
Strain profiles for a lens-shaped InAs quantum dot 
with diameter 15~nm and height 6~nm, embedded in 3~nm and 15~nm thick GaAs buffers.
The hydrostatic strain component $(\epsilon_{xx} +\epsilon_{yy} +\epsilon_{zz})/3$
is plotted with respect to atomic position along the growth direction from
the substrate to the capping layer. The periodic boundary condition is imposed 
on the buffer surface. The simulation with the small buffer underestimates
the compressive strain inside the dot by 0.005 
in comparison with the simulation with the large buffer. 
Furthermore, the small-buffer simulation predicts a tensile strain in the buffer 
while the large-buffer simulation predicts a compressive strain.}
\end{figure}

\begin{figure}[t]\scalebox{0.65}{\includegraphics*{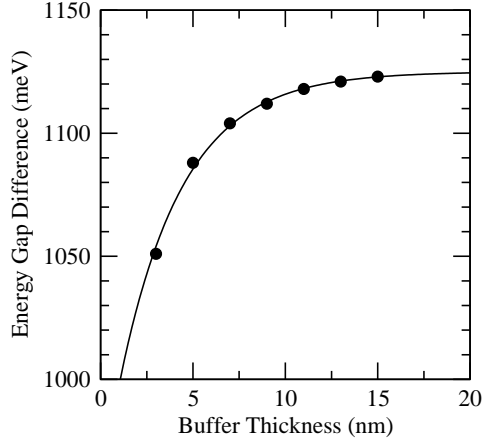}}
\caption{ \label{fig:strain_buffer}
Energy gap between the ground electron and
hole states with respect to untruncated GaAs buffer thickness.
The modeled system is an InAs dot with diameter 15~nm and height 6~nm,
embedded in a GaAs buffer. Both strain profile and electronic structure are calculated with 
the periodic boundary condition imposed on an untruncated buffer surface.
The solid circle is the calculation result, and the line is an exponential fit. 
As the buffer thickness increases and the strain in the dot saturates,
the energy gap converges to 1.125~eV.}
\end{figure}

\begin{figure}[t]\scalebox{0.65}{\includegraphics*{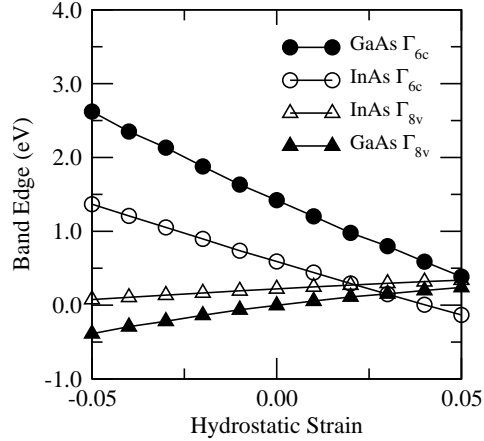}} 
\caption{ \label{fig:strain_bulk} 
Conduction and valence band edges at $\Gamma$ with respect to hydrostatic
strain for bulk InAs and GaAs. The compressive strain increases the direct 
band gap while the tensile strain decreases the gap.} 
\end{figure}

\subsection{Long-ranged Strain Field}\label{sec:strain}
An accurate strain profile is a prerequisite for the 
electronic-structure calculation because the strain field strongly affects
ionic potentials and thus changes the electron Hamiltonian.
In order to obtain an accurate strain profile of InAs/GaAs 
nanostructures, a sufficiently large GaAs buffer needs to be included 
in the simulation domain.
Figure~\ref{fig:strain_profile} shows the dramatic difference between the strain 
profiles calculated with a 3~nm thick buffer and a 15~nm thick buffer. 
The simulation with the small buffer underestimates the compressive 
strain inside the dot and misrepresents the strain in the buffer.
The simulation with the large buffer yields the relaxation of strain at the buffer surface. 
The result indicates that the 15~nm thick buffer is sufficiently large to 
accommodate the strain relaxation that would occur in a realistically sized system.

The saturation of the strain profile can be also monitored by examining 
the convergence of the resulting electronic structure.
Figure~\ref{fig:strain_buffer} shows the energy gap between the ground electron 
and hole states with respect to the buffer size used for both 
strain and electronic structure calculations.  
Both the strain profile and the electronic structure are calculated  
with the periodic boundary condition.    
As the buffer thickness varies from 3~nm to 15~nm,
the resulting energy gap increases by about 72~meV (from 1.051~eV to 1.123~eV).
The large gap change demonstrates the long-range effect of the strain field 
on the electronic structure.
The exponential fit suggests the convergence of the gap to 1.125~eV 
as the buffer thickness becomes infinite.
Since the small buffer underestimates the strain inside the dot, 
the increase of the buffer thickness results in the increase of the dot strain.
Under the compressive hydrostatic strain, the bulk GaAs and InAs conduction
(valence) band edge at $\Gamma$ shifts up (down),
as shown in Figure~\ref{fig:strain_bulk}.
Following the trends, the lowest conduction (the highest valence) electron 
energy of the strained nanostructure increases (decreases) as the buffer 
thickness increases and the dot strain becomes stronger. 
These shifts of the electron energies lead to the overall increase of the
energy gap. Figures~\ref{fig:strain_profile} and \ref{fig:strain_buffer} clearly
demonstrate the importance of a sufficiently large buffer size 
in the simulation domain in order to obtain both accurate strain profile 
and electronic structure.  

Although the strain calculation requires a large buffer,
an accurate electronic structure can be obtained with a smaller buffer due 
to the finite extent of the localized electron wave functions. 
Using a truncated buffer will ease the
computational requirements for the electronic structure calculation since 
the dimension of the Hamiltonian grows linearly with the number of atoms
included in the model.  From here on,
the electronic structure is calculated with a truncated buffer while 
keeping the equilibrium atomic positions obtained from the strain calculation
using a 15~nm thick buffer and implementing the boundary conditions 
addressed in Section~\ref{sec:boundary}.
The efficiency and reliability of each boundary condition 
are systematically analyzed in terms of the elimination 
of non-physical surface states in Section~\ref{sec:surface},
the number of Lanczos iterations required for interior-state energy convergence in
Section~\ref{sec:convergence}, the insensitivity of the converged energy to the boundary
energy shift in Section~\ref{sec:shift}, and the buffer size required 
for the energy convergence in Section~\ref{sec:buffer}.

\begin{figure}[t]\scalebox{0.6}{\includegraphics*{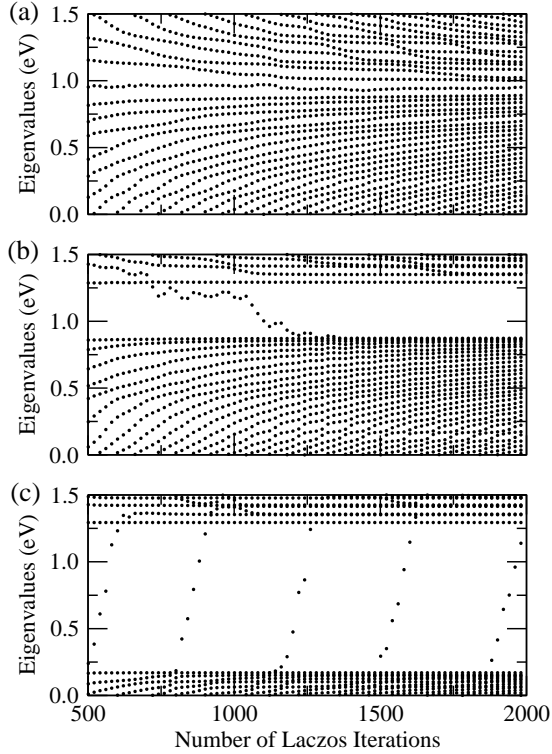}}
\caption{ \label{fig:eigenvalue} 
Eigenvalues of the Lanczos tridiagonal matrix versus the number of
Lanczos iterations (a) without any modification to boundary energies,
(b) with the boundary condition of raising surface-atom orbital energies (BC~I),
and (c) with the boundary condition of raising dangling-bond energies (BC~II).
The modeled system is an InAs dot
with diameter 15~nm and height 6~nm, embedded in GaAs. The strain is calculated
with a 15~nm thick GaAs buffer, while the electronic structure is calculated
with a truncated buffer with thickness 3~nm. The energy shifts for the boundary
condition are set to be (b) $\delta_s$=5~eV, $\delta_p$=3~eV, and
(c) $\delta_{sp^3}$=5~eV.} 
\end{figure}

\subsection{Surface/Interface State Elimination}\label{sec:surface}
One important criterion for a proper BC is the elimination of non-physical 
surface/interface states from the energy region of interest.
Figure~\ref{fig:eigenvalue} presents the eigenvalues obtained from the Lanczos
iterations when three different boundary conditions are applied to a 3~nm thick truncated
buffer. First, to visualize the importance of having a proper boundary condition, 
the eigenvalues without any modification to the boundary energies are plotted in
Figure~\ref{fig:eigenvalue}(a). When such a trivial boundary condition is
implemented, many surface states are formed, which prevents the Lanczos
algorithm from resolving eigenvalues for the physical interior states. 
By comparison, Figure~\ref{fig:eigenvalue}(b) and (c) show that
BC~I and II remove surface states and develop an energy gap. 
The energy shifts used in this calculation are $\delta_s$=5 eV,
$\delta_p$=3 eV, and $\delta_{sp^3}$=5 eV. BC~II efficiently eliminates all
non-physical surface states in the middle of the gap between about 0.3 eV and
1.2 eV.  In contrast, BC~I does not remove all the surface states. 
The dense spectrum of the remaining surface states prevents the convergence
of bound hole states below 0.3~eV. 

\begin{figure}[t] \scalebox{0.6}{\includegraphics*{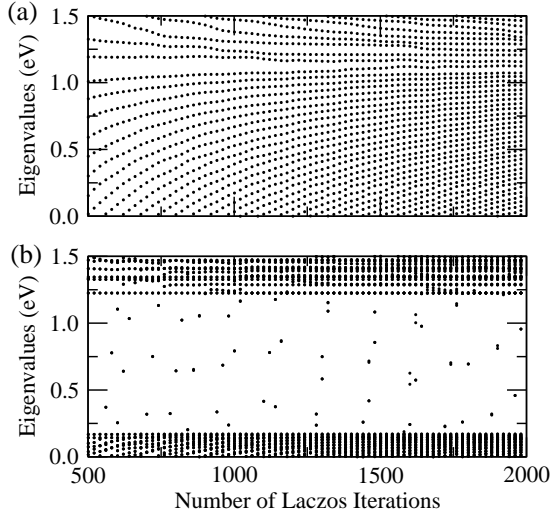}}
\caption{ \label{fig:period} 
Eigenvalues of the Lanczos tridiagonal matrix versus the number of Lanczos iterations 
with the periodic boundary condition (BC~III) (a) using the truncated buffer with thickness 
3~nm and (b) using the untruncated buffer with thickness 3~nm. 
The difference between the two buffers lies in the equilibrium positions of atoms, since 
the former buffer uses the result of the strain calculation with a 15~nm thick buffer while
the latter buffer uses that with a 3~nm thick buffer. The strain profile results for the two
cases are shown in Fig.~\ref{fig:strain_profile}.}
\end{figure}

BC~III is also applied to the truncated buffer to test its efficiency in 
interface-state elimination. 
Figure~\ref{fig:period} shows the eigenvalues of the Lanczos tridiagonal matrix
with the periodic boundary condition:  (a) using a truncated buffer
with thickness 3~nm and (b) using an untruncated buffer with thickness 3~nm. 
In the former the strain profile is calculated with a 15~nm thick buffer
and then the buffer is reduced to 3~nm to calculate the electronic structure, 
while in the latter both the strain profile and electronic structure 
are calculated with a 3~nm thick buffer. 
In both cases, the periodic boundary condition is imposed for not only the electronic 
structure calculation but also the strain profile. 
The periodic boundary condition with the truncated buffer results in many
spurious states in the middle of the gap, while that with the untruncated
buffer does not.  

\begin{figure}[t]\scalebox{0.45}{\includegraphics*{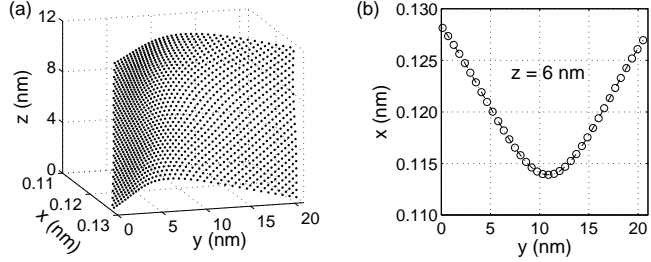}} 
\caption{ \label{fig:interface} 
Atomic positions at the boundary plane of the truncated GaAs buffer: (a) 3-D
visualization of the boundary plane, (b) a slice through a plane with z about
6~nm. The plane is bent due to non-uniform strain generated by the lattice
mismatch between the InAs dot and GaAs buffer. The variation of the
atomic positions along the $x$ axis is about 5\% of the unstrained bond length
of 0.24~nm.} 
\end{figure}

The mid-gap states in the truncated-buffer simulation are formed because of 
the non-planar interface at the boundaries.  
A lattice mismatch of 7\% between InAs and GaAs induces
strain in both the InAs dot and the GaAs buffer. The strain bends the boundary
plane of the truncated buffer by as much as 5\% of the unstrained GaAs bond length
(see Fig.~\ref{fig:interface}). When the bent boundaries are
connected by the periodic boundary condition, the bond between the atoms at the
interface is significantly stretched or compressed. The strained bonds result
in non-physical ``interface'' states in the middle of the gap.
As shown in Figure~\ref{fig:strain_bulk}, strain dramatically change
the band structure of bulk GaAs --- 
tensile strain reduces the band gap while compressive strain increases the gap. 
Similarly, the strongly strained interface in the truncated buffer yield mid-gap states. 
In contrast, the boundaries of the untruncated buffer 
are smooth due to the periodic boundary condition imposed on the strain calculation. 
As a result, it does not yield interface states. 
However, because of its inaccurate strain profile the resulting electronic structure 
is also inaccurate as discussed in Section~\ref{sec:strain}.  

To avoid the unrealistic interface states induced by the truncated periodic BC,
the atomic positions of the truncated buffer need to be adjusted to flatten the
interface.\cite{wang-kim-zunger} However, the adjustment unavoidably leads to 
an inaccurate strain profile unless the truncated buffer is large enough for strain to
saturate near the interface. We have experimented with a partial relaxation of
the boundary layers but found unsatisfactory results --- many interface states
remain, because the partial relaxation is not sufficient to flatten the
interface. To succeed in eliminating interface states, one should start with a 
larger buffer whose boundary is less strained so that the partial relaxation
can lead to a flat boundary. 
 
BC I and II do not require any adjustment to the interface of the truncated buffer, 
as opposed to BC~III which requires an artificial flattening of the interface. 
Therefore, we conclude that the non-periodic BCs are more efficient than the 
periodic BC in terms of the elimination of surface or interface states 
with a smaller truncated buffer while accurately incorporating 
the strain profile resulting from a larger-buffer simulation.  

\begin{table}[t]
\caption{\label{tab:lanczos}
Number of Lanczos iterations required to obtain
eigenvalues converged within 0.1~$\mu$eV 
with the boundary condition of raising orbital energies of
surface atoms (BC~I) and with the boundary condition of raising dangling-bond
energies (BC~II). The modeled system is a lens-shaped InAs quantum dot with a
diameter 15~nm and height 6~nm, embedded in a 3~nm thick GaAs buffer. 
The strain profile is obtained with a 15~nm thick GaAa buffer.} 
\begin{ruledtabular}
\begin{tabular}{ccc}
No. of eigenvalues & BC I & BC II \\
\hline
1 & 1250 & 650 \\
2 &2320 & 1370 \\
3 & 2400 & 1370 \\
4 & 2420 & 1370
\end{tabular}
\end{ruledtabular}
\end{table}

\subsection{Eigenvalue Convergence Speed}\label{sec:convergence}

To investigate the efficiencies of BC~I and II in resolving interior-state
energies, the speed of the eigenvalue convergence is measured in terms of the number
of Lanczos iterations required. Table~\ref{tab:lanczos} lists the number of
Lanczos iterations required for a given number of converged eigenvalues for
BC~I and II.  BC~II results in a faster convergence than BC~I. For example, to
acquire four eigenvalues, BC~II requires half as many iterations as BC~I. The
efficiency of BC~II is attributed to the elimination of the dense spectrum of
surface states. In general, iterative eigenvalue solvers easily find
eigenvalues in a sparse spectrum, but show difficulty resolving eigenvalues in
a dense spectrum. Therefore, the search of interior states is accelerated by the
elimination of surface states from the interior-state spectrum.

\begin{figure}[t]\scalebox{0.65}{\includegraphics*{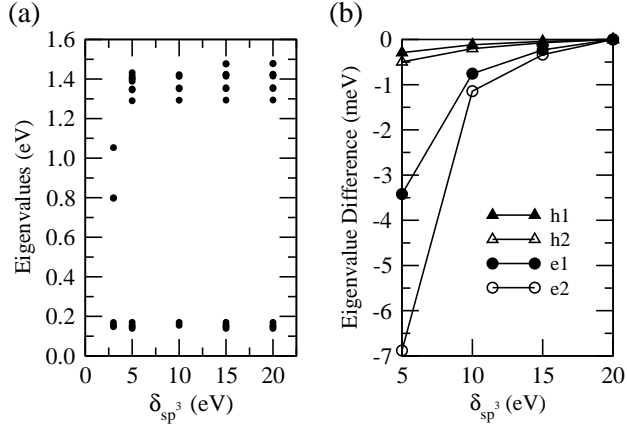}} 
\caption{ \label{fig:shift} 
(a) Electron energy versus dangling-bond energy shift $\delta_{sp^3}$,
(b) Variations of the ground and excited electron (e1, e2) and
hole (h1, h2) energies with respect to energy shift.
A energy shift larger than 5~eV eliminates surface
states in the middle of the gap between 0.2 and 1.2~eV. The electron and hole
energies vary only by a few meV when the energy shift varies from 5 to 20~eV.}
\end{figure}

\begin{figure}[t]\scalebox{0.6}{\includegraphics*{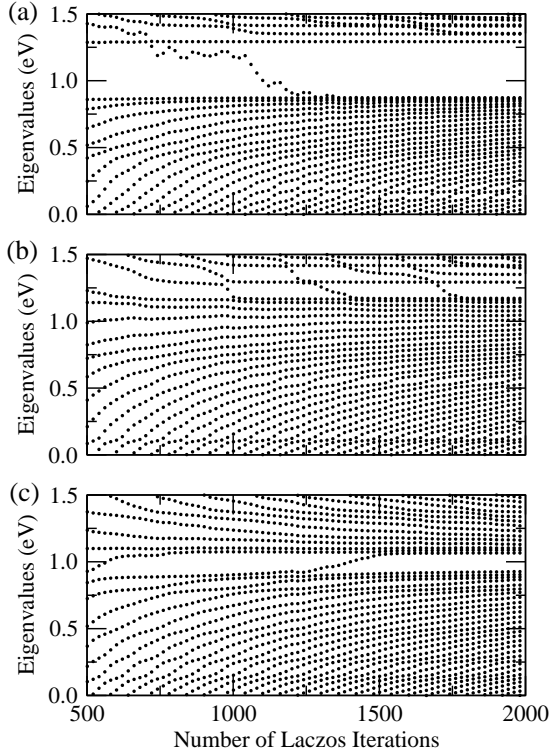}}
\caption{\label{fig:orbital}
Eigenvalues of the Lanczos tridiagonal matrix versus the number of
Lanczos iterations with the boundary condition of raising orbital energies
(BC I), (a) using $\delta_s$=5~eV and $\delta_p$=3~eV, (b) using $\delta_s$=5~eV
and $\delta_p$=4~eV, and (c) using $\delta_s$=20~eV and $\delta_p$=20~eV.
We have not found any pair of $\delta_s$ and $\delta_p$ that
succeeds in removing all the surface states in the
middle of the gap which is between 0.2 and 1.2~eV.} 
\end{figure}

\subsection{Boundary Energy Shift}\label{sec:shift}

To implement BC~I and II, appropriate boundary energy shifts $\delta_s$, $\delta_p$, 
and $\delta_{sp^3}$ must be determined.  The ultimate goal in choosing the energy
shift is to eliminate all surface states in the energy region of interest for
interior states (e.g., within the band gap). Figure~\ref{fig:shift} shows
converged eigenvalues with respect to the energy shift $\delta_{sp^3}$ in BC~II. 
While $\delta_{sp^3}$=3 eV leads to surface states in the middle of the gap,
the energy shift larger than 5 eV eliminates all the surface states and 
leads to the eigenvalues converged within a few meV. This
indicates that the electronic structure is insensitive to the choice of the
energy shift in BC~II if the shift is big enough to remove all surface states. 

In contrast, the effect of energy shifts on the electronic structure with BC~I
is highly unpredictable; a slight change of the shifts leads to a completely
different Lanczos eigenvalue spectrum. For instance, changing $\delta_p$ from
3~eV to 4~eV results in more surface states within the gap, as shown in
Figure~\ref{fig:orbital}.  A wide range of positive and negative 
energy shifts $\delta_s$ and $\delta_p$ was tested to achieve the 
best performance for eliminating surface
states. However, no pair of tested $\delta_s$ and $\delta_p$ within 20~eV
succeeded in eliminating all the surface states and in yielding the band gap
1.1~eV which is given by both BC~II with a truncated buffer (see
Fig.~\ref{fig:eigenvalue}~(c)) and BC~III with an untruncated buffer (see
Fig.~\ref{fig:period}~(b)).  
This inefficiency in removing surface states is attributed to the
truncation of connected bonds. BC~I truncates both dangling bonds and connected
bonds, while BC~II truncates only the dangling bonds. Since the connected bond
should be connected to interior atoms, the truncation of the connected bond
will create a dangling bond to the interior atoms, and the dangling bond gives
rise to surface states within the gap. This result suggests that BC I has
intrinsic difficulties in removing surface states. 
 
\begin{figure}[t]\scalebox{0.65}{\includegraphics*{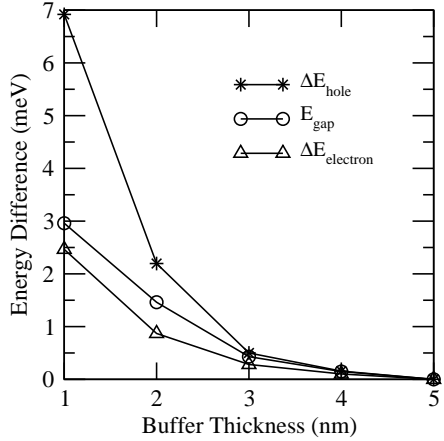}} 
\caption{ \label{fig:buffer_size} 
Variations of the energy gap (${\rm E}_{\rm gap}$)
between the ground electron and hole states,
the energy spacing ($\Delta {\rm E}_{\rm electron}$)
between the ground and the first excited electron states, and
the energy spacing ($\Delta {\rm E}_{\rm hole}$)
between the ground and the first excited hole states,
with respect to the truncated buffer thickness
for an InAs quantum dot with diameter 15~nm and height 6~nm.
The boundary condition of raising dangling-bonds energies 
(BC~II) with $\delta_{sp^3}$=10~eV is used for this calculation.
${\rm E}_{\rm gap}$, $\Delta {\rm E}_{\rm electron}$, and $\Delta {\rm
E}_{\rm hole}$ with each buffer thickness are subtracted by those with
buffer thickness 5~nm to obtain the variations of these quantities.
As the buffer thickness becomes larger than 3~nm,
${\rm E}_{\rm gap}$, $\Delta {\rm E}_{\rm electron}$, and $\Delta {\rm
E}_{\rm hole}$ converge to 1123~meV, 56~meV, and 14~meV within 1~meV,
respectively.}
\end{figure}

\subsection{Buffer Size}\label{sec:buffer}

To find a reasonable buffer size for accurate electronic-structure
calculations, the quantitative dependence of the electronic structure on the
buffer size is examined. BC~II is used since it provides
the most efficient elimination of non-physical states.  Figure~\ref{fig:buffer_size}
presents the energy gap between the lowest conduction electron and the highest
valence electron levels for different buffer thicknesses. The buffer thickness
is defined as the distance between the faces of the buffer GaAs box and the
InAs dot. When the buffer thickness is bigger than 3~nm, 
the energy gap and the electron and hole energy spacings converge
to 1123~meV, 56~meV, and 14~meV within 1~meV, respectively. 
This convergence indicates that a 3~nm thick buffer is large enough 
to obtain the electronic structure with the accuracy of 1~meV.
In general, the optimal buffer size varies with quantum-dot
size and electron level, and hence one should determine the optimal size by
monitoring the convergence of the energies for a desired accuracy. 

\section{Conclusions}

In summary, we have investigated three types of boundary conditions for the
electronic structure of a self-assembled InAs dot embedded in GaAs within the 
framework of the empirical tight-binding model. 
Two non-periodic boundary conditions
demonstrate higher efficiency than the truncated periodic boundary condition,
in terms of the buffer size required to eliminate non-physical mid-gap states. 
Between the non-periodic boundary conditions, BC~II (raising dangling-bond
energies) more efficiently removes surface states than BC~I (raising orbital
energies of surface atoms). Therefore, BC~II is identified as the most
efficient boundary condition for eliminating surface states and 
achieving the convergence of interior-state energies with a truncated buffer. 

The effect of the dangling-bond energy shift and the buffer size on the
electronic structure have been further examined with the efficient BC~II.  An
energy shift bigger than 5~eV efficiently removes all spurious states 
in the middle of the gap, and yields an energy gap insensitive to the
further increase of the energy shift. For a lens-shaped InAs dot with diameter 
15~nm and height 6~nm, the GaAs buffer thickness of 3~nm is large enough  
to obtain the electronic structure with the accuracy of 1~meV. 

While our new boundary condition (BC~II) has been developed within the framework of
empirical tight binding, it can be extended to other models. An example is to use
an empirical pseudo-potential with a non-local part that is a sum of projections on
sub-spaces with well-defined orbital momentum.\cite{williamson-wang-zunger} 
In this case, a transformation of the basis set to the $sp^3$ hybridized orbitals 
can be performed and an energy shift can be applied solely to the dangling 
bonds as presented in this work. 

Boundary condition II with a truncated buffer takes advantage of the localization 
of the electron wave functions in a core nanostructure 
such as the InAs/GaAs quantum dot illustrated
in this article. This scheme is not straightforwardly applicable to other types 
of heterostructures where electrons or holes are localized in the buffer. 
However, if the core nanostructure is larger than the extent of the 
electron or hole wave function localized in the buffer, one can truncate the core
region instead of the buffer. 
When one of the carriers (electron or hole) is localized in the core 
and the other carrier in the buffer, the core carrier can be 
modeled with a truncated buffer and the buffer carrier with a truncated core, 
so long as the coupling between the conduction 
and valence bands is weak enough to treat the electron and hole Hamiltonians independently. 

\begin{acknowledgments} 
This work was performed at Jet Propulsion Laboratory,
California Institute of Technology under a contract with the National
Aeronautics and Space Administration. This work was supported by grants from
NSA/ARDA, ONR, and JPL internal Research and Development. 
\end{acknowledgments}


\begin{thebibliography}{19}
\expandafter\ifx\csname natexlab\endcsname\relax\def\natexlab#1{#1}\fi
\expandafter\ifx\csname bibnamefont\endcsname\relax
  \def\bibnamefont#1{#1}\fi
\expandafter\ifx\csname bibfnamefont\endcsname\relax
  \def\bibfnamefont#1{#1}\fi
\expandafter\ifx\csname citenamefont\endcsname\relax
  \def\citenamefont#1{#1}\fi
\expandafter\ifx\csname url\endcsname\relax
  \def\url#1{\texttt{#1}}\fi
\expandafter\ifx\csname urlprefix\endcsname\relax\def\urlprefix{URL }\fi
\providecommand{\bibinfo}[2]{#2}
\providecommand{\eprint}[2][]{\url{#2}}

\bibitem[{\citenamefont{Klimeck et~al.}(1995)\citenamefont{Klimeck, Lake,
  Bowen, Frensley, and Moise}}]{klimeck_apl}
\bibinfo{author}{\bibfnamefont{G.}~\bibnamefont{Klimeck}},
  \bibinfo{author}{\bibfnamefont{R.~K.} \bibnamefont{Lake}},
  \bibinfo{author}{\bibfnamefont{R.~C.} \bibnamefont{Bowen}},
  \bibinfo{author}{\bibfnamefont{W.~R.} \bibnamefont{Frensley}},
  \bibnamefont{and} \bibinfo{author}{\bibfnamefont{T.}~\bibnamefont{Moise}},
  \bibinfo{journal}{Appl. Phys. Lett.} \textbf{\bibinfo{volume}{67}},
  \bibinfo{pages}{2539} (\bibinfo{year}{1995}).

\bibitem[{\citenamefont{Mamaluy et~al.}(2003)\citenamefont{Mamaluy, Sabathil,
  and Vogl}}]{mamaluy}
\bibinfo{author}{\bibfnamefont{D.}~\bibnamefont{Mamaluy}},
  \bibinfo{author}{\bibfnamefont{M.}~\bibnamefont{Sabathil}}, \bibnamefont{and}
  \bibinfo{author}{\bibfnamefont{P.}~\bibnamefont{Vogl}}, \bibinfo{journal}{J.
  Appl. Phys.} \textbf{\bibinfo{volume}{93}}, \bibinfo{pages}{4628}
  (\bibinfo{year}{2003}).

\bibitem[{\citenamefont{Klimeck et~al.}(2002)\citenamefont{Klimeck, Oyafuso,
  Boyking, Bowen, and von Allmen}}]{klimeck-nemo3d}
\bibinfo{author}{\bibfnamefont{G.}~\bibnamefont{Klimeck}},
  \bibinfo{author}{\bibfnamefont{F.}~\bibnamefont{Oyafuso}},
  \bibinfo{author}{\bibfnamefont{T.~B.} \bibnamefont{Boyking}},
  \bibinfo{author}{\bibfnamefont{R.~C.} \bibnamefont{Bowen}}, \bibnamefont{and}
  \bibinfo{author}{\bibfnamefont{P.}~\bibnamefont{von Allmen}},
  \bibinfo{journal}{Computer Modeling in Engineering and Science}
  \textbf{\bibinfo{volume}{3}}, \bibinfo{pages}{601} (\bibinfo{year}{2002}).

\bibitem[{\citenamefont{Oyafuso et~al.}(2003)\citenamefont{Oyafuso, Klimeck,
  von Allmen, and Boykin}}]{oyafuso}
\bibinfo{author}{\bibfnamefont{F.}~\bibnamefont{Oyafuso}},
  \bibinfo{author}{\bibfnamefont{G.}~\bibnamefont{Klimeck}},
  \bibinfo{author}{\bibfnamefont{P.}~\bibnamefont{von Allmen}},
  \bibnamefont{and} \bibinfo{author}{\bibfnamefont{T.~B.}
  \bibnamefont{Boykin}}, \bibinfo{journal}{Phys. Status Solidi B}
  \textbf{\bibinfo{volume}{239}}, \bibinfo{pages}{71} (\bibinfo{year}{2003}).

\bibitem[{\citenamefont{Wang et~al.}(1999)\citenamefont{Wang, Kim, and
  Zunger}}]{wang-kim-zunger}
\bibinfo{author}{\bibfnamefont{L.-W.} \bibnamefont{Wang}},
  \bibinfo{author}{\bibfnamefont{J.}~\bibnamefont{Kim}}, \bibnamefont{and}
  \bibinfo{author}{\bibfnamefont{A.}~\bibnamefont{Zunger}},
  \bibinfo{journal}{Phys. Rev. B} \textbf{\bibinfo{volume}{59}},
  \bibinfo{pages}{5678} (\bibinfo{year}{1999}).

\bibitem[{bon()}]{bond_direction}
\bibinfo{note}{When a system is under non-hydrostatic strain, the bond
  directions are different from those in an unstrained structure. However, the
  difference is small enough to ignore in the modeled system, where the maximum
  strain is about 5\%. Therefore, we use $sp$ hybridized orbitals that are
  aligned along the bond directions of the unstrained structure.}

\bibitem[{\citenamefont{Cardona}(1983)}]{cardona}
\bibinfo{author}{\bibfnamefont{M.}~\bibnamefont{Cardona}},
  \bibinfo{journal}{Phys. Status Solidi B} \textbf{\bibinfo{volume}{118}},
  \bibinfo{pages}{463} (\bibinfo{year}{1983}).

\bibitem[{sur()}]{surface_passivation}
\bibinfo{note}{The explicit modeling of surface passivation would lead to a
  larger Hamiltonian dimension and consequently more computation time than
  BC~II due to the passivation atoms.}

\bibitem[{\citenamefont{Jancu et~al.}(1998)\citenamefont{Jancu, Scholz,
  Beltram, and Bassani}}]{jancu}
\bibinfo{author}{\bibfnamefont{J.-M.} \bibnamefont{Jancu}},
  \bibinfo{author}{\bibfnamefont{R.}~\bibnamefont{Scholz}},
  \bibinfo{author}{\bibfnamefont{F.}~\bibnamefont{Beltram}}, \bibnamefont{and}
  \bibinfo{author}{\bibfnamefont{F.}~\bibnamefont{Bassani}},
  \bibinfo{journal}{Phys. Rev. B} \textbf{\bibinfo{volume}{57}},
  \bibinfo{pages}{6493} (\bibinfo{year}{1998}).

\bibitem[{\citenamefont{Schmidt et~al.}(1997)\citenamefont{Schmidt,
  Medeiros-Ribeiro, Garcia, and Petroff}}]{schmidt-petroff}
\bibinfo{author}{\bibfnamefont{K.~H.} \bibnamefont{Schmidt}},
  \bibinfo{author}{\bibfnamefont{G.}~\bibnamefont{Medeiros-Ribeiro}},
  \bibinfo{author}{\bibfnamefont{J.}~\bibnamefont{Garcia}}, \bibnamefont{and}
  \bibinfo{author}{\bibfnamefont{P.~M.} \bibnamefont{Petroff}},
  \bibinfo{journal}{Appl. Phys. Lett.} \textbf{\bibinfo{volume}{70}},
  \bibinfo{pages}{1727} (\bibinfo{year}{1997}).

\bibitem[{\citenamefont{Garcia et~al.}(1997)\citenamefont{Garcia,
  Medeiros-Ribeiro, Schmidt, Ngo, Feng, Lorke, Kotthaus, and
  Petroff}}]{garcia-petroff}
\bibinfo{author}{\bibfnamefont{J.~M.} \bibnamefont{Garcia}},
  \bibinfo{author}{\bibfnamefont{G.}~\bibnamefont{Medeiros-Ribeiro}},
  \bibinfo{author}{\bibfnamefont{K.}~\bibnamefont{Schmidt}},
  \bibinfo{author}{\bibfnamefont{T.}~\bibnamefont{Ngo}},
  \bibinfo{author}{\bibfnamefont{J.~L.} \bibnamefont{Feng}},
  \bibinfo{author}{\bibfnamefont{A.}~\bibnamefont{Lorke}},
  \bibinfo{author}{\bibfnamefont{J.}~\bibnamefont{Kotthaus}}, \bibnamefont{and}
  \bibinfo{author}{\bibfnamefont{P.~M.} \bibnamefont{Petroff}},
  \bibinfo{journal}{Appl. Phys. Lett.} \textbf{\bibinfo{volume}{71}},
  \bibinfo{pages}{2014} (\bibinfo{year}{1997}).

\bibitem[{\citenamefont{Keating}(1966)}]{keating}
\bibinfo{author}{\bibfnamefont{P.}~\bibnamefont{Keating}},
  \bibinfo{journal}{Phys. Rev.} \textbf{\bibinfo{volume}{145}},
  \bibinfo{pages}{637} (\bibinfo{year}{1966}).

\bibitem[{\citenamefont{Pryor et~al.}(1998)\citenamefont{Pryor, Kim, Wang,
  Williamson, and Zunger}}]{pryor}
\bibinfo{author}{\bibfnamefont{C.}~\bibnamefont{Pryor}},
  \bibinfo{author}{\bibfnamefont{J.}~\bibnamefont{Kim}},
  \bibinfo{author}{\bibfnamefont{L.~W.} \bibnamefont{Wang}},
  \bibinfo{author}{\bibfnamefont{A.~J.} \bibnamefont{Williamson}},
  \bibnamefont{and} \bibinfo{author}{\bibfnamefont{A.}~\bibnamefont{Zunger}},
  \bibinfo{journal}{J. of Appl. Phys.} \textbf{\bibinfo{volume}{83}},
  \bibinfo{pages}{2548} (\bibinfo{year}{1998}).

\bibitem[{\citenamefont{Boykin et~al.}(2002)\citenamefont{Boykin, Klimeck,
  Bowen, and Oyafuso}}]{boykin_strain}
\bibinfo{author}{\bibfnamefont{T.~B.} \bibnamefont{Boykin}},
  \bibinfo{author}{\bibfnamefont{G.}~\bibnamefont{Klimeck}},
  \bibinfo{author}{\bibfnamefont{R.~C.} \bibnamefont{Bowen}}, \bibnamefont{and}
  \bibinfo{author}{\bibfnamefont{F.}~\bibnamefont{Oyafuso}},
  \bibinfo{journal}{Phys. Rev. B} \textbf{\bibinfo{volume}{66}},
  \bibinfo{pages}{125207} (\bibinfo{year}{2002}).

\bibitem[{\citenamefont{L\"owdin}(1950)}]{lowdin}
\bibinfo{author}{\bibfnamefont{P.-O.} \bibnamefont{L\"owdin}},
  \bibinfo{journal}{J. Chem. Phys.} \textbf{\bibinfo{volume}{18}},
  \bibinfo{pages}{365} (\bibinfo{year}{1950}).

\bibitem[{\citenamefont{Harrison}(1999)}]{harrison}
\bibinfo{author}{\bibfnamefont{W.~A.} \bibnamefont{Harrison}},
  \emph{\bibinfo{title}{Elementary Electronic Structure}}
  (\bibinfo{publisher}{World Scientific, New Jersey}, \bibinfo{year}{1999}).

\bibitem[{\citenamefont{Slater and Koster}(1954)}]{slater-koster}
\bibinfo{author}{\bibfnamefont{J.~C.} \bibnamefont{Slater}} \bibnamefont{and}
  \bibinfo{author}{\bibfnamefont{G.~F.} \bibnamefont{Koster}},
  \bibinfo{journal}{Phys. Rev.} \textbf{\bibinfo{volume}{94}},
  \bibinfo{pages}{1498} (\bibinfo{year}{1954}).

\bibitem[{\citenamefont{Treffethen and III}(1997)}]{treffethen}
\bibinfo{author}{\bibfnamefont{L.~N.} \bibnamefont{Treffethen}}
  \bibnamefont{and} \bibinfo{author}{\bibfnamefont{D.~B.} \bibnamefont{III}},
  \emph{\bibinfo{title}{Numerical Linear Algebra}} (\bibinfo{publisher}{Society
  for Indistrial and Applied Mathematics, Philadelphia}, \bibinfo{year}{1997}),
  pp. \bibinfo{pages}{276--284}.

\bibitem[{\citenamefont{Williamson et~al.}(2000)\citenamefont{Williamson, Wang,
  and Zunger}}]{williamson-wang-zunger}
\bibinfo{author}{\bibfnamefont{A.~J.} \bibnamefont{Williamson}},
  \bibinfo{author}{\bibfnamefont{L.-W.} \bibnamefont{Wang}}, \bibnamefont{and}
  \bibinfo{author}{\bibfnamefont{A.}~\bibnamefont{Zunger}},
  \bibinfo{journal}{Phys. Rev. B} \textbf{\bibinfo{volume}{62}},
  \bibinfo{pages}{12963} (\bibinfo{year}{2000}).

\end{thebibliography}
\end{document}